\documentclass[aps,prb,floats,twocolumn,showpacs]{revtex4}

\usepackage{epsfig}

\usepackage{amsmath}
\usepackage[dvips]{color}
\usepackage{amssymb}

\usepackage{bm}

\begin{document}

\title{Shot noise in NS junctions with Weyl superconductor}

\author{ A. Golub}
\affiliation{Debpartment of Physics, Ben-Gurion University of
  the Negev Beer-Sheva, Israel}
\date{\today}

\begin{abstract}
We demonstrate that current-current correlations (in particular the shot noise),  can be used to study the intrinsic superconductivity in a slightly doped Weyl semi-metal. The systems studied is an N-WS
tunneling junction where the  left electrode is a normal metal while the right electrode is a Weyl superconductor (WS).
 The superconductivity supports  surface state with crossed flat bands thereby impact the low energy spectrum.
 This spectrum displays a modified density of states in the gap region that strongly affects transport characteristics of the N-WS junction. The Fano factor is calculated as function of the applied bias, and  shown to be dependent essentially on the orientation  of the surface of WS relative to  the tunneling direction. If this  orientation supports the occurrence of low energy state, then the shot noise power decreases with decreasing voltage, a property similar to that prevailing in a junction with Majorana bound state.

\end{abstract}


\maketitle
\noindent
 \section{ Introduction}
   Occurrence of new  class of materials, referred to as Weyl semimetals (WSM),  was predicted
theoretically \cite{vish1,ran,balent,dai5,teo1,dai4,dai3,fang2,hasan6} and recently realized in experiments \cite{hasan5,fang,hasan4,fu,hasan,dai,cava,dai2,hasan9,hasan3,sato2,kharzeev,wang}.
The superconductivity of Weyl semimetals (WSM) were also studied theoretically \cite{moore,moore2,moore3,aji,zuzin,meng,gorio,sato} and
  experimentally \cite{sun}. In the normal state
 WSMs have, in general, a finite bulk conductance and hence they can be considered as metallic. At the same time, a semi-metal phase is realized when the
Fermi energy touches the Weyl point where
there is a  contact between the filled valence and empty conduction bands.
 Interestingly, the Weyl points act as monopole source in momentum space for Berry phase.

  Semi-metallicity is unstable against doping that moves the Fermi level into conduction or valence bands, thereby
 reaching the phase of Weyl metal or even a WS. WSM host Fermi arcs and require breaking
 time-reversal or inversion symmetries.
Weyl metals preserve
 the topological properties of Fermi surface like Fermi arcs. These arcs
 are terminated by projections of Weyl points on the surface Brillouin zone.
  The superconductivity of doped WSMs  supports crossed flat bands and not only  simple arcs\cite{sato}.

 \noindent
In this work we expose the physics of an N-WS junction between a normal metal (N) and a
 WS with uniform pairing state (which is essential for the occurrence of crossed flat bands).
 The novel aspect here, of course, is encoded in the superconducting side of the junction,
 that is remarkably distinct from that of a usual superconductor. Indeed,
 upon slight doping with finite chemical potential $\mu$, a WSM has disconnected Fermi
surfaces, each of which surrounds a band-touching
Weyl points. (Doping of WSM is easily
achieved  by modifying the chemical potential such that
it is much larger than the gap of the WS, that is, $\mu \gg \Delta$.)
A special form of the order parameter  can be realized on these Fermi surfaces,
corresponding either to BCS s-wave phase or  to the Fulde-Ferrell-
Larkin-Ovchinnikov (FFLO) phase.
In the first case, the pairing occurs between states related by inversion symmetry(if it exists), while in the FFLO phase the states of opposite sides of each Fermi surface  are paired.
A constant s-wave
pairing can support bulk gap
nodes on the Fermi surface \cite{moore} (nodal superconductor).

To elucidate the peculiar physics of the N-WS junction we
write down the corresponding  low energy Hamiltonian, and then calculate the shot noise power of
the junction and display it as a function of applied voltage.
 It was shown \cite{sato} that
  depending on the orientation of the surface of the WS relative to  tunneling direction the  novel surface states  can strongly affect conductance. Here we consider the impact of these states on the  shot noise power.

 Technically, our approach uses the Green's functions (GFs) of WS integrated over momentum. The method describes point junctions and is similar to the approach undertaken in paper \cite{arnold}.
 The GF depends crucially on the boundary conditions at the N-WS surface of contact,
 and therefore encodes the corresponding direction of the tunneling electrons. It is
three cases are considered here: (1-2) N-WS junction
with respective tunneling directions along the x (N-WSx) and z axes (N-WSz), and 3)   (for comparison), N-S
junction between normal metal and homogenous s-wave superconductor.
\noindent
\section{ Model}
 The low energy Hamiltonian for the N-WS junction reads $H=H_N+H_T+H_{WS}$, where the left lead (N) is biased by voltage $V$ and its Hamiltonian reads $H_N=\frac{1}{2}\sum_{k}\epsilon_kc^{\dagger}_{kL}\sigma_0 \tau_z c_{kL}$. The tunneling Hamiltonian for the point contact is $ H_T=wc^{\dagger}_{L}(0)\sigma_0 \tau_z c_{ R}(0)+h.c$, where $w$ is the  junction tunneling constant. On the right side, we describe the WS
by low energy quasiclassical Hamiltonian \cite{moore,aji,meng,sato}
 \begin{eqnarray}\label{HSW}
    H_{SW}&=&\frac{1}{2}\sum_{k}c^{\dagger}_{kR}[v(k_x\sigma_y \tau_z-k_y\sigma_x \tau_0)-v_z k_z\sigma_z \tau_z \nonumber\\
    && -\mu\sigma_0 \tau_z -\Delta\sigma_y \tau_y]c_{kR}
 \end{eqnarray}
where $v_z=t_z\sin(Q)$, $k_i=-i\partial_i$, ($i=x,y,z$) and $v$ ($t_z$) is the hopping in the $k_x,k_y$ plane (along the $k_z$ axis).
The Hamiltonian (\ref{HSW}) encodes the low energy spectrum near the Weyl point (0,0,Q). This  spectrum plays a crucial role for study of transport and noise processes. Denoting
$c_{\bf k}$=$(c_{k\uparrow},c_{-k\downarrow},c^{\dagger}_{-k\uparrow},c^{\dagger}_{k\downarrow})^{T}$
 we can write
\begin{eqnarray}
 H_{SW} &=& \tfrac{1}{2} \sum_{\bf k} c_{\bf k}^\dagger {\cal H}_{\bf k} c_{\bf k} \label{H}
\end{eqnarray}
Then ${\cal H}_{\bf k}$  corresponds to  a $4 \times 4$ operator valued matrix acting in spin$\otimes$Nambu space
in which  $\sigma_i$( $\tau_i$) are the respective Pauli matrices.
A useful technique to properly handle the boundary conditions at the interface
is to use the Hamiltonian (\ref{HSW}) also for the N side ($x<0$). For this we
set $\Delta(x<0)=0$ and add to $H_{SW}$ a term $M\sigma_z\tau_z \theta(-x)$ \cite{vish3,sato}.
At the end we take the limit $M \rightarrow \infty$. The Hamiltonian  which is correct also beyond the low energy regime is given by
 a minimal two-band model Hamiltonian \cite{moore,aji,meng,sato}
 \begin{eqnarray}
   {\cal H}_{\bf k} &=& v\sin{k_x}\sigma_y\tau_z-v\sin{k_y}\sigma_x\tau_0+(v_z\cos{k_z}-M_0)\sigma_z\tau_z\nonumber\\
   && +m(2-\cos{k_x}-\cos{k_y})\sigma_z\tau_z-\mu\sigma_0\tau_z-\Delta\sigma_y\tau_y
 \end{eqnarray}
 where $M_0=t_z\cos{Q}$ denotes a parameter (like magnetic order) that breaks the time-reversal symmetry. By changing the other parameter $m$ we can change the number and positions  of Weyl points.

The current operator is defined as derivative of number of electrons in the left lead (say):
\begin{equation}\label{j}
    j=-ie[N_L,H]=-ie[\frac{w}{2}c^{\dagger}_L(0)\sigma_0 \tau_0c_R(0)-h.c].
\end{equation}
To proceed, we write down the corresponding Keldysh action to which
we add a source field $\alpha(t)$ that multiplies
the  current operator. Explicitly,
\begin{eqnarray}
  S_{ef} &=& \sum_{k,k'}\int dt \{ \mbox{Tr} [c^{\dagger}_k\hat{g}^{-1}_{k,k'}c_{k' }] \}. \label{action} \\
  \hat{g}^{-1}_{k,k'} &=&  g^{-1}_{k}\delta_{k,k'}-\Sigma_{T,k,k'}, \nonumber\\
   \Sigma_{T,k,k'}&=&w A_{k,k'}\sigma_0(\varrho_x \tau_z\rho_0+i\varrho_y\alpha\tau_0\rho_x).
\end{eqnarray}
Here the subscript $k=({\bf k} L), ({\bf k} R)$ of the spinors $c_k$ refers also to (L,R) (left,right) space described by $\varrho$ matrices  and to Keldysh space described by $\rho$ matrices.
Moreover,  $A_{k,k'}=1$ presents a constant matrix in momentum space $k,k'$.
The Green functions (GF) of the leads are diagonal in LR space, that is,  $g^{-1}_k= g^{-1}_{Lk}P_{\varrho+}+g^{-1}_{Rk}P_{\varrho-}$ where $P_{\varrho \pm}=(1\pm \varrho_z)/2$. We denote
 $g_{Lk}$ as the Kedysh GF of normal metal and $g_{Rk}$
as GF of the superconductor in NS junction. The crucial point is that $g_{Rk}$ depends on
the direction of the  tunneling electrons,
which, in turns, affects the shot noise and the conductance of the N-WS junction.

Now we are in a position to write down the expressions for current and current noise following variation of the action (\ref{action}) with respect to the quantum source field $\alpha$
\begin{eqnarray}
  J &=& \frac{ie}{2} \mbox{Tr} [\hat{g} \frac{\partial}{\partial \alpha(t)}\hat{g}^{-1}]\\
  S(t,t') &=& \frac{-e^2}{4} \mbox{Tr} [\hat{g}( \frac{\partial}{\partial \alpha(t)}\hat{g}^{-1})\hat{g} \frac{\partial}{\partial \alpha(t')}\hat{g}^{-1}]
\end{eqnarray}
where trace includes also the integration  over momentum and time variables.
Expressions for the GF $\hat{g}$ are obtained by calculating the inverse of block matrix $\hat{g}^{-1}$ in LR space. After this is completed, summation over momentum in the expressions for the current and the noise  can be easily carried out. The results are presented solely in terms of GF integrated over momentum, given by,
\begin{eqnarray}
\bar{G}=\left(
                   \begin{array}{cc}
                                                                 G_{LL} & G_{LR} \\
                                                                 G_{RL} & G_{RR} \\
                                                               \end{array}
                                                             \right),
\end{eqnarray}
where $G_{ii}=2\pi N_i g_{ii}$ and $G_{ij}=\frac{4x}{w}g_{ij}$  ($i\neq j$). Here $g_{ii}= (\bar{g}_{i}^{-1}-4x\tau_z\bar{g}_{j} \tau_z)^{-1}$,
$g_{ij}=\bar{g}_i\tau_zg_{jj}$.
The effective tunneling width $x=\pi^2 w^2N_L N_W$ depends on the density of states of
the normal metal lead $N_L$ as well as on the density of states of the WS
$N_{W}=\mu^2/(4\pi^2 v^2 v_z)$.
Thus, the shot noise power and the stationary current in terms of these GFs acquire a form
\begin{eqnarray}
  J &=& \frac{ew}{2} \mbox{Tr}[i\hat{g}\varrho_y\rho_x]=2e x \mbox{Tr}(g^K_{LR}-g^K_{RL}), \label{cur}\\
 S &=& -e^2 x (S_1-4x S_2), \label{noise1}\\
  S_1 &=& 2 \mbox{Tr} [g_{LL}\rho_x g_{RR}\rho_x],\nonumber\\
  S_2&=& \mbox{Tr} [g_{LR}\rho_x g_{LR}\rho_x +(L\leftrightarrows R)],
\end{eqnarray}
where the superscript $K $ denotes the Keldysh  GF. Explicit expressions for Keldysh, retarded and advanced GF, as well as for the noise and the tunneling current, are presented in the  Supplementary Material (appendix).

\noindent
\section{Shot noise in NS junctions}
\subsection { N-S junction with s-wave superconductor}
The homogeneous s-wave superconductor has no states in the gap region. The
imaginary part of diagonal component of $g_R^r$ integrated over momentum  is equal to zero in the gap region.
After the momentum integration the matrix GF  $g_R^r$ for s-wave pairing has a  form
\begin{eqnarray}
  g^r(\nu) &=&-i [\alpha^r(\nu)\tau_0+\beta^r(\nu)\tau_x]\sigma_0 \\
  \alpha^r(\nu) &=& \frac{|\nu|\Theta(|\nu|-1)}{\sqrt{\nu^2 -1}}+
  \frac{\nu \Theta(-|\nu|+1)}{i\sqrt{-\nu^2 +1}}\nonumber
\end{eqnarray}
and $\beta^r(\nu)=\alpha^r(\nu)/\nu$ (see \cite{arnold}). Here $\nu=\epsilon/\Delta$, the step-function $Q(y)=1$ if $y>0$ and is equal to zero if $y<0$.
 We calculate shot noise and Fano factor, the tunneling current and conductance for this type of N-S junction.
  The result as function of tunneling transparency are presented by panels (b) of  Figs.1-3. The conductance in the all figures is given in terms of transmission coefficient : $T_n=4x/(1+x)^2$. Also here and below we consider only the zero temperature limit.

   The conductance for N-S junctions
is similar to the  one obtained by BTK \cite{btk} (compare our Fig.3(b) with Fig.7 in reference  \cite{btk}). As to the Fano factor, we can see the resemblance of our Fig.1(b) with Fig.3 of the work \cite{datta}.

\subsection { N-WSz junction, tunneling along $\hat {\bf z}$}
As we have already noted, the GFs for the N-WS junctions are determined by boundary conditions at the contact surface,  and the result strongly depends on the respective tunneling direction \cite{sato}. For tunneling
 along the $z$ direction, an approximate expression for the GF follows directly from the Hamiltonian (1),
  in which all derivatives are considered as momenta (including the $z$-component).
This approximation is justified for tunneling in $z$ direction
 because the surface of the WS that is perpendicular to the $z$-axis does  not introduce new low energy states. The effective order parameter of WS is equal to zero value at the poles of Fermi surface. To see this let us at first,
for notational convenience,  renormalize  the components of the momentum as $k_{x,y}/(\mu/v)\rightarrow k_{x,y}$, $k_z /(\mu/v_z)\rightarrow k_z $.   This enables us to write\\ $ g_{Rk}^r
 $=$\frac{1}{\mu}[\frac{\epsilon}{\mu} \sigma_0\tau_0$-$k_x\sigma_y \tau_z$+$k_y\sigma_x \tau_0)$+$k_z\sigma_z \tau_z $+$\sigma_0 \tau_z$+$\frac{\Delta}{\mu} \sigma_y \tau_y]^{-1}$. \\
Inverting this $4 \times 4$ matrix involves a denominator which contains a  factor $\epsilon^2-\xi^2-(\Delta \sin\theta)^2$ where, in rescaled form, $\xi =(|\vec{k}|-1)$.
 The principal ingredient for evaluating conductance and noise is the GF integrated over momentum.
 The denominator which appears in $g_{Rk}^r$  (whose detailed expression is given in Supplementary Material), clearly indicates a nodal structure of the effective order parameter.

\noindent
\subsection{ N-WSz junction  tunneling along $\hat {\bf z}$: Results} 
 The  panels (a)
 of Figs. 1-3 display the Fano factor, current density and conductance.
 The results show a  similarity with those pertaining to d-wave superconductors \cite{tanaka,tanaka2,tanaka3,ting}.

 Indeed, the approach which we use correlates with that in references \cite{btk,datta,tanaka2,tanaka3}. We have shown in the case  N-S junctions haw our method is related to some early works. As to N-WSz junction,  the conductance presented by our Fig.3(a)
 is similar to Fig.2 of reference \cite{tanaka2}, whereas Fano factor shows some enhancement around zero voltage (our Fig.1(a) similar to that of reference \cite{tanaka3}. Also at voltages $eV<1.5\Delta$ the Fano factor behaves like one for junction with d-wave superconductor (see Fig.2(c) in the reference \cite{ting}).
 \begin{figure} [!ht]
\centering
\includegraphics [width=0.4 \textwidth]{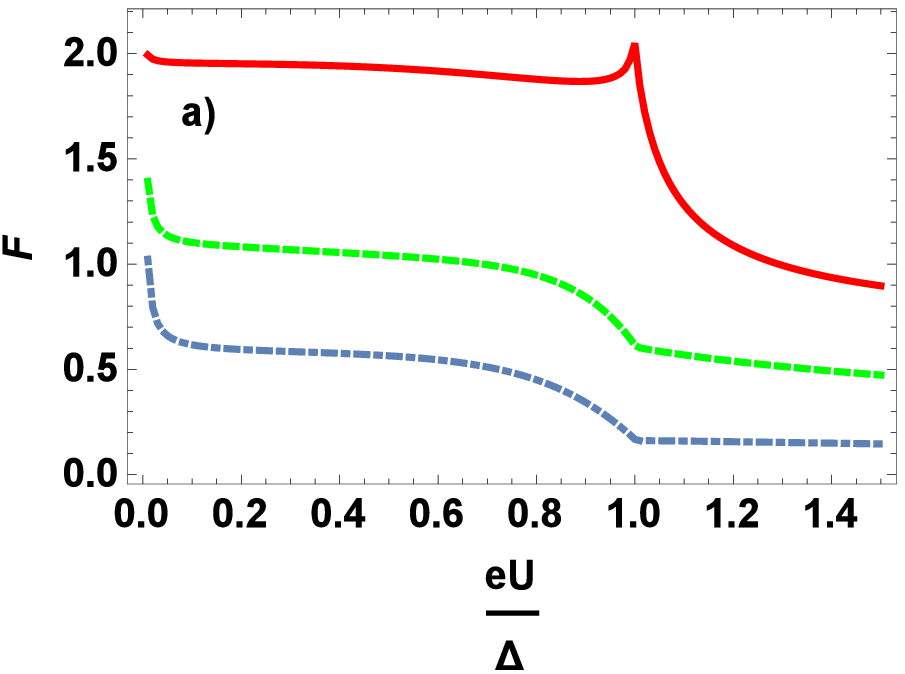}
\includegraphics [width=0.4 \textwidth]{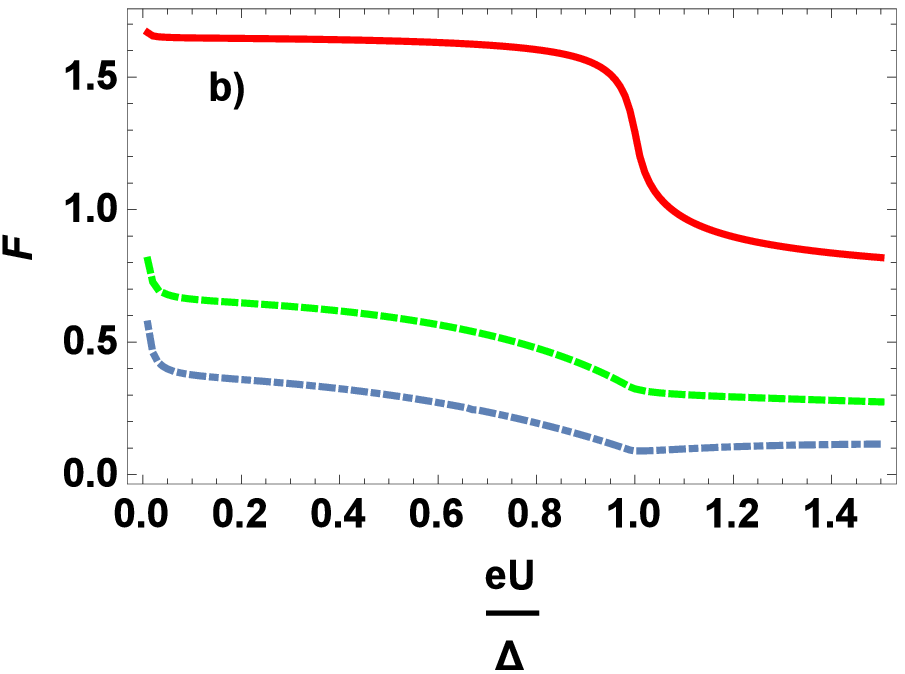}
\caption {\footnotesize{(Color online) Fano factor as function of applied voltage $eU/\Delta$  for different values of tunneling width $x$. Solid(red), green ( dashed) and dot-dashed curves
correspond to x=0.1, 0.5 and 0.7,respectively.  panel a) presents N-WSz junction,  panel b) stands for N-S contact where $S$ denotes a standard gapped s-wave  superconductor.
 }}
\label{Fig1}
\end{figure}
\begin{figure} [!ht]
\centering
\includegraphics [width=0.4 \textwidth]{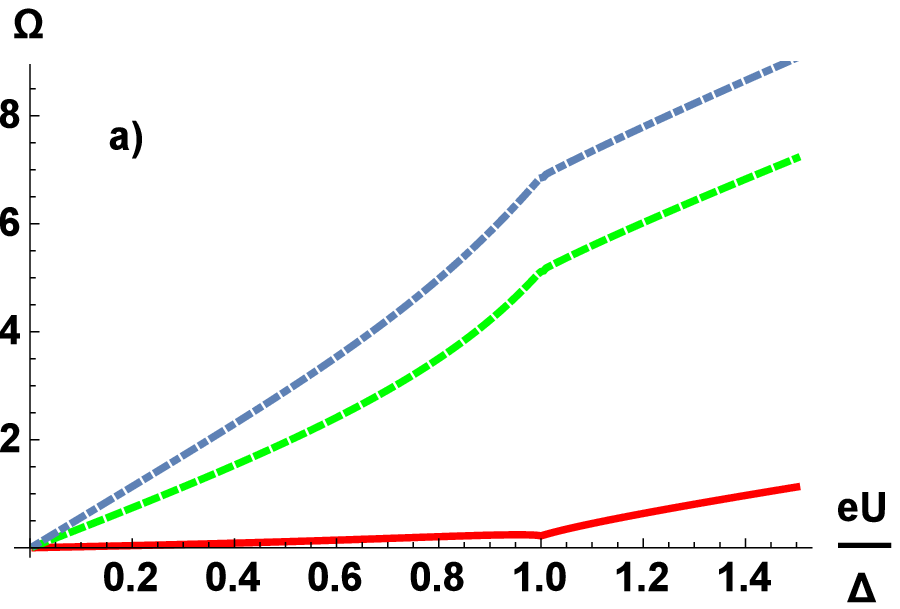}
\includegraphics [width=0.4 \textwidth]{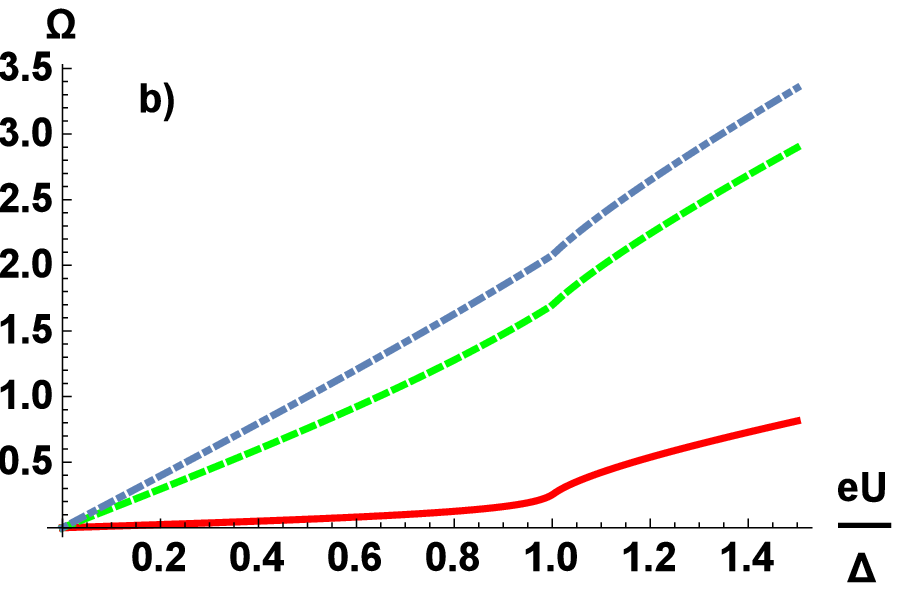}
\caption {\footnotesize{Color online) The same junctions as in Fig.1.( a) and b)) The plot represents the current ($\Omega=J/e$) -voltage dependence.}}
\label{Fig2}
\end{figure}
\begin{figure} [!ht]
\centering
\includegraphics [width=0.4 \textwidth]{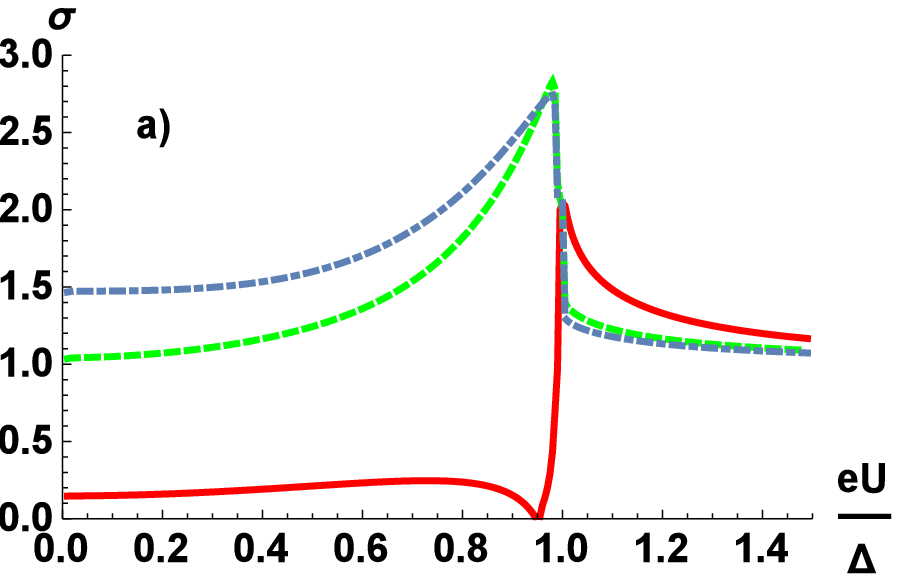}
\includegraphics [width=0.4 \textwidth]{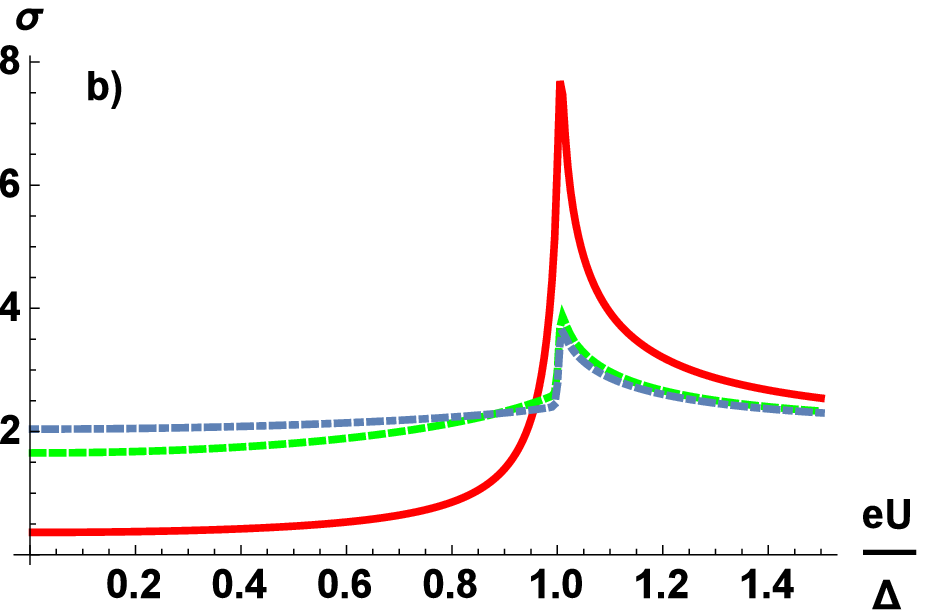}
\caption {\footnotesize{(Color online) The conductance $ \sigma$ (normalized by normal conductance) as function of applied voltage with the same set of parameters and the same junctions as in Fig.1.( a) and b))
 }}
\label{Fig3}
\end{figure}

\begin{figure} [!ht]
\centering
\includegraphics [width=0.4 \textwidth]{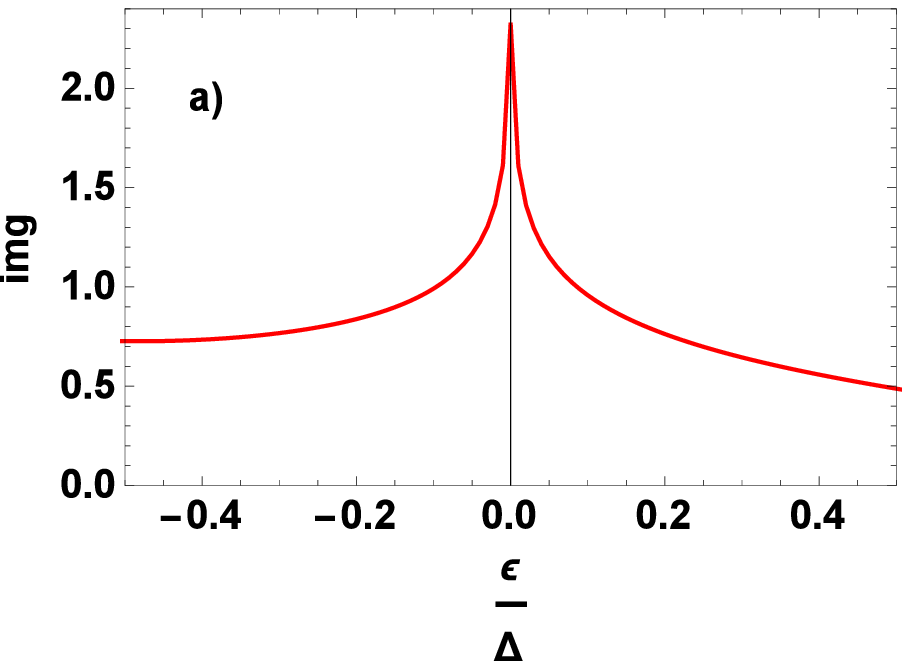}
\includegraphics [width=0.4 \textwidth]{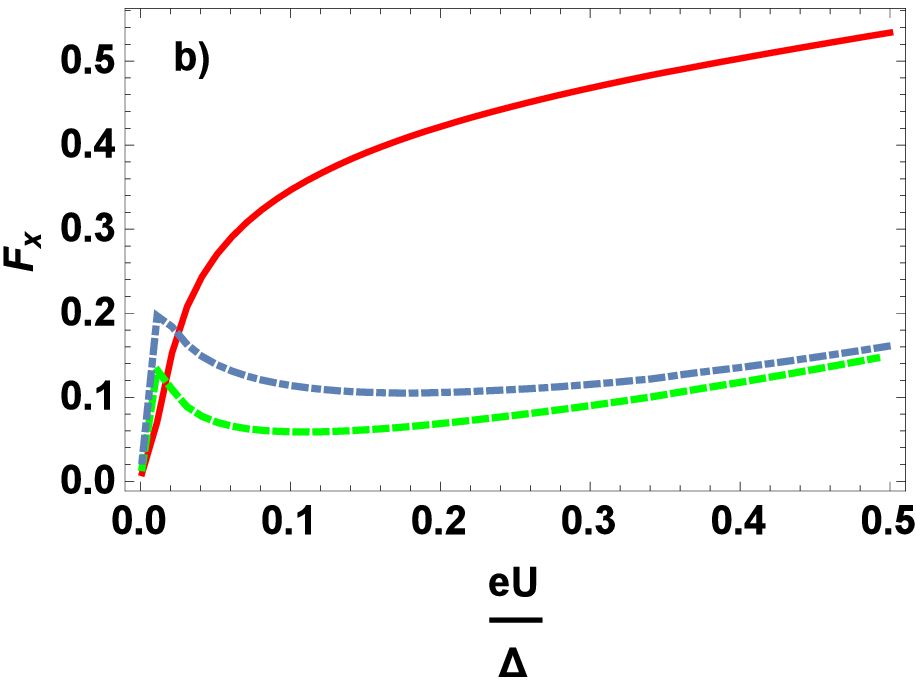}

\caption {\footnotesize{(Color online)   a) Density of states for N-WSx junction versus energy in the gap region. A peak of the density of states is  related to topological reconstruction of the spectra. The panel b) presents Fano factor  of N-WSx junction as function of  applied voltage. The  parameters for tunneling in x-direction are the same as in Fig.1. The  lines correspond to x=0.1 (red solid), 0.5 (green dashed), 0.7 (dot-dashed).
 }}
\label{Fig4}
\end{figure}
 \begin{figure} [!ht]
\centering
\includegraphics [width=0.4 \textwidth]{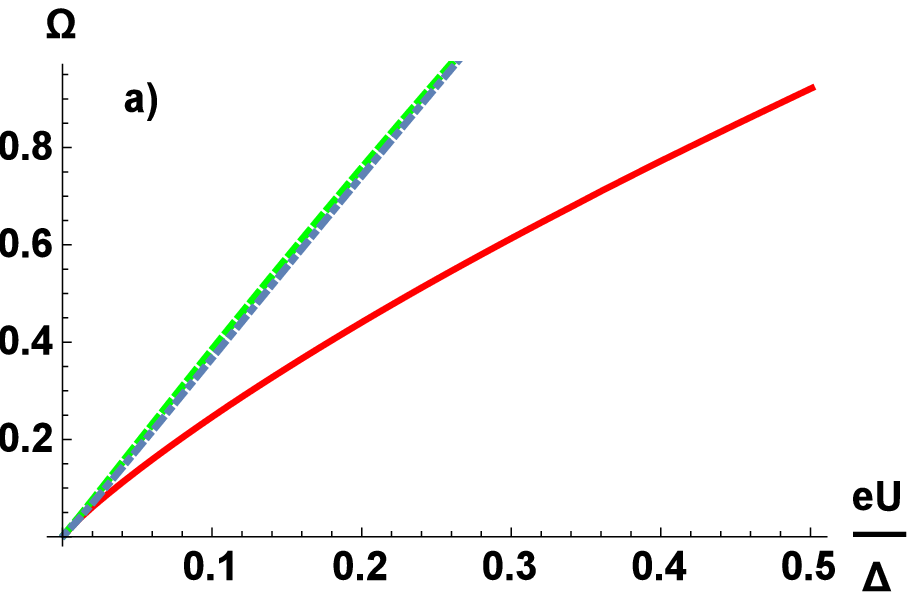}
\includegraphics [width=0.4 \textwidth]{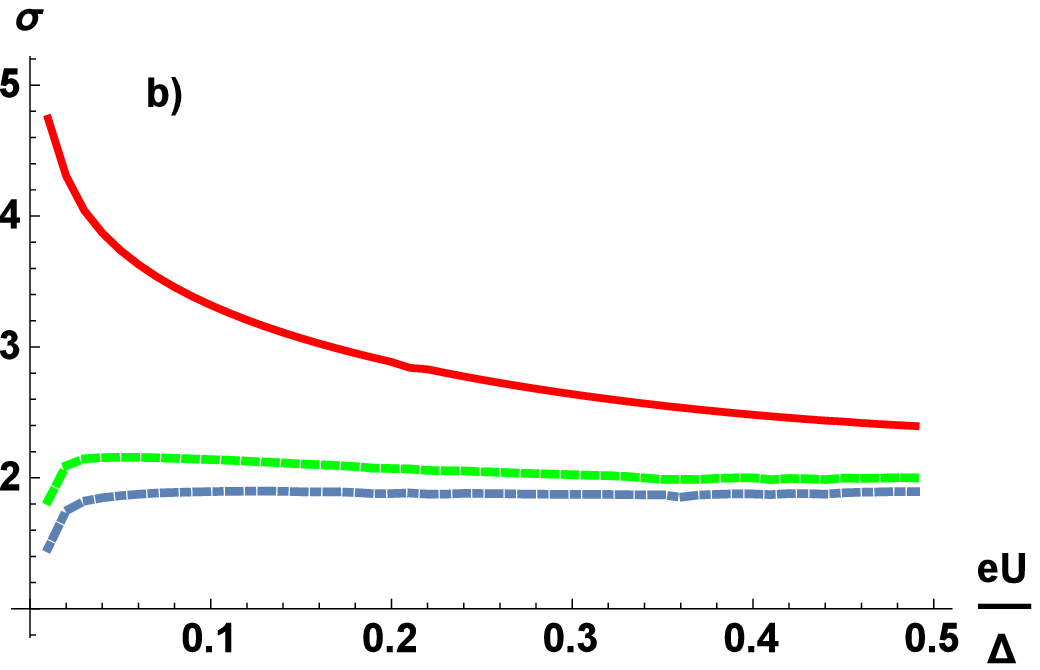}
\caption {\footnotesize{(color online)  a) The I-V characteristics of  N-WSx junction   and b) conductance  in x-direction as function of voltage. The  parameters are the same as in Fig.4 (b).
 }}
\label{Fig5}
\end{figure}

\noindent
\subsection { N-WSx junction, tunneling along $\hat {\bf x} $}
The GF of the WS satisfies boundary conditions on a surface perpendicular to the x-axis (at x=0).  This is the basic element required for the derivation of the shot noise and electron transport along $\hat {\bf x}$, where
the topological nature of the WS is clearly exposed \cite{sato}. The behavior of the conductance and the shot noise
is determined by the structure of
the flat band and the nature of the low energy states that occur
 for  $\epsilon\ll \Delta$. We must obtain Green function  $g_R^r$ to find the spectrum and calculate the shot noise power.
The expression for the corresponding GF (detailed derivation is given in Supplementary Material) reads
 \begin{equation}\label{gx}
  g_R^r = g_d\hat{I}-g_{off} \tau_y\sigma_y
 \end{equation}
 where $g_d$ and $g_{off}$ are integrated over momentum the diagonal and the off-diagonal  components of $g_R^r$.

 \subsection { N-WSx junction, tunneling along $\hat {\bf x} $: Results} 
 The  imaginary part $g_R^r$  (diagonal components)  displays a peak of the density of states
in the middle of superconductive gap region (Fig.4, a) ).

The  panel b) of Fig.4 presents the Fano factor, for the N-WSx junction
(with tunneling along the x- direction), while
 the tunneling current and the conductance for this case are, respectively, displayed on the a)  and b) panels of Fig.5.
 At the smallest value of transmission parameter  $x=0.1$ there is a peak  of conductance at $V\rightarrow0$ which is similar to the zero bias conductance peak obtained in reference \cite{sato} for nonzero control parameter  $m$.
 Indeed a weak zero-bias conductance peak is maintained till $x=0.4$ (not shown in the Fig.5). It is not completely clear why at $x>0.4$ a weak conductance peak is slightly shifted from $V=0$. However, we note that the zero-bias conductance peak is a distinctive characteristic (at least for $x<0.4$) of electron transport for tunneling in the x-the direction of N-WSx. There is no such an effect  for ordinary N-S junction ( Fig.3 b)) or for N-WSz junction with tunneling in z-direction ( Fig.3 a)).

 As far as the shot noise, Fano factor are concerned,
comparing our results for  usual N-S  and  N-WSz junctions with  N-WSx junctions we note a remarkable feature.
  Unlike the cases with S-wave superconductor (Fig.1 b)) and N-WSz junction (Fig1. a) ), the Fano factor of N-WSx junction   (Fig.4 b)) is strongly reduced at $ V->0$ reaching zero value  at $V=0$. The shot noise  power itself  tends to zero even faster. Thus the  N-WSx junctions have unique properties encoded in the conductance  and  shot noise voltage dependence. Therefore, the shot noise together with conductance  can help to test experimentally
the superconductivity in a doped Weyl semi-metal. The effect  has simple explanation: The flat band low energy
spectrum \cite{sato} which arises in N-WSx junction causes the occurrence of the peak in the density of states  at zero energy (in the gap region) (Fig.4  panel a)) which  is responsible
for the zero-bias peak of conductance and strong reducing of the shot noise at $V\rightarrow 0$.

Here we  would like to compare our system with N-MBS junction \cite{me,haim,fazio,trip}, where MBS stand for topological superconductor with  Majorana zero bound state at his end. In both systems there is zero energy bound state (in the gap region). We expect behavior of the shot noise similar to our system. Indeed,  the formula for the total shot
 noise power in the case of Majorana zero bound state \cite{me} is given by equation
\begin{eqnarray}
  S &=& \frac{8e^2\Gamma}{h}(\arctan\frac{eV}{2\Gamma}-\frac{2eV\Gamma}{(eV)^2+4\Gamma^2})
\end{eqnarray}
where $\Gamma$ is the tunneling width.
Thus we see that as the bias voltage $V\rightarrow 0$  {\it both the noise power and its first derivative} vanish: $(\frac{dS}{dV})_{V=0}=0$. i. e. the transport is coherent. Moreover, it is found that the conductance has zero bias peak.


\noindent
\section{ Conclusions} In this work we suggest that measurements  of shot noise and Fano factor serve as an additional
benchmark  for studying the topological properties of Weyl superconductors and the superconductivity of a doped Weyl semi-metal. The superconductivity itself can support crossed surface flat bands \cite{sato}. The specific topology of WSM is presented by point nodes in the s-wave pairing state. For the model described by the Hamiltonian (1), different tunneling directions are not equivalent: for z direction tunneling, the boundary does not creates in-gap states, whereas for transport along the x axis the flat bands define the low energy spectrum of Weyl superconductor. These low energy bands   cause the finite density of states in the middle of the superconducting gap. The high density of states in the gap strongly influences the transport and the shot noise power. The Fano factor tends to zero at $V\rightarrow 0$-the effect that can be tested on experiment.

In closing we remark on the approximations that were made here. First, the parameter $m$ which controls the positions and number of the Weyl points is not employed in our
calculations while it was considered for complete two-band model (3) in reference \cite{sato}.
However, based on the approximate quasiclassical Hamiltonian (1), the nontrivial topology related to the surface of WS orthogonal to the x-axis clearly reveals itself via  zero bias peak of the conductance and the vanishing at $V\rightarrow 0$ of both the shot noise power and the Fano factor.
Second,  we used an approach that describes the electron tunneling in the same way
as for a point junction. The right electrode is
a superconductor which is characterized by its GFs  properly  integrated over momenta. However, while for a z- directed transport the  inhomogeneity along the z-axis is irrelevant and may be ignored in deriving the shot noise, for x-tunneling direction the inhomogeneity along the x-axis (due to the superconductor's surface)  is relevant. In the latter case we have solved the boundary problem and found an approximate expression for the GF.

\noindent
\begin{acknowledgments}
I would like to thank B. Horovitz, Y. Avishai  and E. Grossfeld     for stimulating discussions and  for valuable remarks.
\end{acknowledgments}
.



\noindent
\begin{acknowledgments}
I would like to thank B. Horovitz, Y. Avishai  and E. Grossfeld  for stimulating discussions and  for valuable remarks.
\end{acknowledgments}

\appendix

\section{Supplementary Material}
\subsection{ Shot noise}
Here we provide the derivation of the shot noise power for  NS junctions. We obtain a general expression for shot noise which is correct for ordinary s-wave and for Weyl superconductors. We also  present the total list of Green's functions which were used.

The noise formula in the main text (Eqs 11,12) after taking the trace in Keldysh space acquires a form
\begin{eqnarray}
  S_1 &=& 2tr\{4Im[g^r_{LL}] Im[g^r_{RR}]+g^K_{RR}g^K_{LL}\}\\
  S_2&=& tr\{4Im[g^r_{LR}] Im[g^r_{LR}]+g^K_{LR}g^K_{LR}+(L\leftrightarrows R)\}\nonumber
\end{eqnarray}
 The trace also includes integration over energy variable $\nu$.
The GFs which are involved in above expression for noise have a form:

(a)The  left (N) electrode GFs without tunneling interaction (we have dropped here and below the bar at $g$.)
\begin{eqnarray}
  g_L^{K}&=& -i(\tanh\frac{\nu-eV}{2T}P_+ +\tanh\frac{\nu+eV}{2T}P_- )
 \nonumber \\
   g_L^{r,a} &=& \mp\frac{i}{2}\hat{I},\,\,\ P_{\pm} = \frac{1}{2}(\hat{I}\pm\tau_z \sigma_0 )\nonumber
\end{eqnarray}
where $\hat{I}$ is unit matrix in four dimensional case. The superscripts $r$, $a$, $K$ stand for retarded, advanced and Keldysh component of $g$.

(b) The integrated over momentum non-interacting GF of superconductor $g_R$,  as explained in the main text, depends  on orientation of WS surface relative to tunneling direction.
The Keldysh component of this GF is simply $ g_R^K=\tanh[\frac{\nu}{2T}](g_R^r-g_R^a)$.

The left and right electrode GFs are modified by tunneling
\begin{eqnarray}
  g_{LL}^{r,a} &=& \mp \frac{i}{2}(\hat{I}\pm2ix\tau_z\sigma_0 g_R^{r,a}\tau_z \sigma_0)^{-1}\\
  g_{RR}^{r,a} &=& (g_R^{r,a-1}\pm 2ix)^{-1}
\end{eqnarray}
and for Keldysh GFs we have
\begin{eqnarray}
  g_{LL}^{K} &=& \tanh\frac{\nu}{2T}( g_{LL}^{r}- g_{LL}^{a})+4 A_L \nonumber\\
 g_{RR}^{K} &=& \tanh\frac{\nu}{2T}( g_{RR}^{r}- g_{RR}^{a})+4x A_R\nonumber
\end{eqnarray}
where we used notation
\begin{equation}\label{A}
    A_i=g_{ii}^{r}\hat{F} g_{ii}^{a},\,\,\,\hat{F}= g_L^K+i\hat{I} \tanh\frac{\nu}{2T}
\end{equation}
and $i=L,R$.

The crossed GFs  appear due to tunneling possesses. They can be written  in terms of modified by tunneling diagonal GFs of each electrode (A2,A3)
\begin{eqnarray}
  g_{LR}^{r,a} &=& \mp \frac{i}{2}\tau_z\sigma_0 g_{RR}^{r,a}\nonumber\\
  g_{RL}^{r,a} &=& g_R^{r,a}\tau_z\sigma_0 g_{LL}^{r,a}
\end{eqnarray}
The crossed Keldysh GFs acquire a form
\begin{eqnarray}
  g_{LR}^{K} &=& \tanh\frac{\nu}{2T}( g_{LR}^{r}- g_{LR}^{a})+\tau_z\sigma_0 g_R^{r-1} A_R \nonumber\\
 g_{RL}^{K} &=& \tanh\frac{\nu}{2T}( g_{RL}^{r}- g_{RL}^{a})+4g_R^r \tau_z\sigma_0 A_L\nonumber\\
\end{eqnarray}

Using these formulas the current and the shot noise term $S_1$  acquire a form
\begin{eqnarray}
  J &=& 2e x tr\{\hat{F} [g_{RR}^a \tau_z\sigma_0 g_R^{r-1}g_{RR}^r-4 g_{LL}^{a}g_R^r\tau_z\sigma_0 g_{LL}^{r}]\}\nonumber\\
  S_1 &=& 8 tr\{\tanh\frac{\nu}{2T}[( g_{RR}^{r}- g_{RR}^{a})A_L+x A_R (g_{LL}^{r}- g_{LL}^a)]+\nonumber\\
  &&4x A_R A_L\}+S_{1N}
\end{eqnarray}
A more  complicated expression follows  for $S_2 $

\begin{eqnarray}
S_2 &=& 4x tr\{\tau_z\sigma_0 g_R^{r-1}A_R\tau_z\sigma_0 g_R^{r-1}A_R+\nonumber\\
&&16 g_R^{r}\tau_z\sigma_0 A_L g_R^{r}\tau_z\sigma_0 A_L+\nonumber\\
&&\tanh\frac{\nu}{2T}[2( g_{LR}^{r}- g_{LR}^{a})\tau_z\sigma_0 g_{R}^{r-1}A_R+\nonumber\\
  &&8 (g_{RL}^{r}- g_{RL}^a)g_R^r\tau_z\sigma_0 A_L ]+ S_{2N}
\end{eqnarray}
 Here $S_{1N}$, $S_{2N}$ present the  Naikwist part of the noise power, and are  given, respectively,  by the fist terms in $ S_1$, $S_2$ (A1) with a factor
$(1-(\tanh\frac{\nu}{2T})^2)$.
\subsection{Integrated Green's Functions}
 The 3D  momentum integration of $g_{Rk}^{r}$ for z-tunneling direction consists of  integrations on  $\xi$ and on polar angles ($\theta,\varphi$)
\begin{eqnarray}
  g_R^r &=& h_1\Theta[1-\nu^2]+h_2\Theta[-1+\nu^2] \nonumber \\
  h_1 &=& x_1 \nu \hat {I} - x_2\tau_y\sigma_y \nonumber\\
  h_2&=& y_1 |\nu| \hat {I}- y_2sign[\nu]\tau_y\sigma_y\nonumber
\end{eqnarray}
here $\nu=\epsilon/\Delta$,  $ \Theta[y]$ is the step function  and
\begin{eqnarray}
  x_1 &=& \frac{1}{4}(\pi - 2 i arcsinh [\frac{\nu}{\sqrt{1 - \nu^2}}] sign[\nu]) \nonumber\\
  x_2 &=& \frac{1}{8\nu}(\pi \nu (1 + \nu^2) +
   2 i |\nu| (v - \nonumber\\
   &&(1 + \nu^2)\arcsin(\frac{\nu}{\sqrt{1 - \nu^2}})) \nonumber\\
   y_1&=&-\frac{i}{4} \ln[(1 + |\nu|)/(-1 + |\nu|)]\nonumber\\
   y_2&=&\frac{1}{8} i (2 |\nu| + (1 + \nu^2) \ln[1 - 2/(1 + |\nu|)])\nonumber
\end{eqnarray}

It is more difficult, though standard,  to obtain the integrated over momentum GF of a superconductor with surface plane perpendicular to x axis. Looking for eigenvalues of $det[g_{R,k}^{-1}]=0$ we find 4 eigenvalues $\pm p, \pm p^*$ and  corresponding four eigenvectors $w_1,w_2,w_3,w_4$
\begin{eqnarray}
  w_1& =& \exp(-i p x)\{\gamma^*\tan\frac{\theta}{2},-i\gamma^*\exp(-i\varphi),\nonumber\\
  && i\exp(-i\varphi)\tan\frac{\theta}{2}, 1\}\nonumber\\
   w_2& =& \exp(i p x)\{\gamma^*\tan\frac{\theta}{2},i\gamma^*\exp(i\varphi),\nonumber\\
  && -i\exp(i\varphi)\tan\frac{\theta}{2}, 1\}\\
   w_3& =& \exp(-ip^*x)\{\gamma\tan\frac{\theta}{2},-i\gamma\exp(-i\varphi),\nonumber\\
 &&  i\exp(-i\varphi)\tan\frac{\theta}{2}, 1\} \nonumber \\
   w_4& =& \exp(ip^*x)\{\gamma\tan\frac{\theta}{2},i\gamma\exp(i\varphi),
   -i\exp(i\varphi)\tan\frac{\theta}{2}, 1\} \nonumber
\end{eqnarray}
here $k_{y}=\sin\theta \sin \varphi$, $k_z=\cos\theta$, $\gamma = \sqrt\frac{\nu - i\zeta}{\nu + i\zeta}$; and $\zeta=\sqrt{\sin^2\theta-\nu^2}$. For eigenvalue $p$ we get
\begin{equation}\label{p}
 p=\sqrt{1-k_x^2-k_y^2-2i\zeta}
\end{equation}

The eigenvalues and their eigenvectors can be obtained also for left (normal metal) electrode.
However, to calculate the GF of superconductor only two eigenvectors of normal metal are sufficient. After taking  the limit $M\rightarrow\infty$ at $x=0$ these vectors acquire a form:  $u_1=\{0,0-1,1\}$, $ u_2=\{-1,1,0,0\}$.

The retarded GF of superconductor satisfies equation $(\nu -H_{SWk})g_R^r(x,x') =\delta(x-x')$ where $H_{SWk}$ is given by expression in square brackets of Eq.(1) (main text) with rescaled values of momentums. The GF can be expressed in terms of eigenvectors. The first column of matrix GF (which is defined by the above (A9,A10) eigenvectors and eigenvalues) can be written as
\begin{eqnarray}
  g_R^r(x,x') &=& \Theta(x-x')[b_1(x')w_1(x)+b_2(x')w_4(x)]+\nonumber \\
  && \Theta(x'-x)[a_1(x')w_1(x)+a_2(x')w_2(x)+\nonumber\\
  &&a_3(x')w_3(x)+a_4(x')w_4(x)]
\end{eqnarray}
where we have took in consideration the convergence of GF at $x\rightarrow \infty$. To find functions $a_i(x)$ and $b_j(x)$
we use boundary conditions at $x=0$ and at $x=x'$. Two terms $a_i$ (i=2,3) are completely defined  by conditions at $ x=x'$.  They are not related with boundary at $x=0$. Therefore, these terms do not contribute in relevant low energy physics and we neglect them in GF. 

 After integration over $\varphi$ and adding contribution of other columns, the matrix GF of Weyl  superconductor acquires a simple form with only two different coefficients:
  \begin{equation}\label{gx}
  g_R^r (\theta) = G_d\hat{I}-G_{off} \tau_y\sigma_y
 \end{equation}
Here
\begin{eqnarray}
  G_d(x=0,\theta) &=&  -\frac{\pi \sin^2\theta d_0 \tan[\frac{\theta}{2}]}{\sqrt{\cos^2\theta+\nu^2}}  \\
  G_{off}(x=0,\theta) &=& -\frac{\pi \nu \sin\theta  d_0 }{\sqrt{\cos^2\theta+\nu^2}}
\end{eqnarray}
were  $\sin\theta$  from $d^3p$ differential has been included in Eqs.(A13,A14). Factor $d_0$ reads
\begin{eqnarray}
  d_0 &=& i[\frac{\Theta(\sin\theta-|\nu|)}{\sqrt{\sin^2\theta-\nu^2}}-\nonumber\\
  &&\frac{2sign(\nu)\Theta(-\sin\theta+|\nu|)}{\pi\sqrt{-\sin^2\theta+\nu^2}}\ln(\sqrt{-1+\beta_-^2}+|\beta_-|)]- \nonumber\\
  &&\frac{2 sign(\nu)  \Theta(\sin\theta-|\nu|)}{\pi\sqrt{\sin^2\theta-\nu^2}}\ln(\sqrt{1+\beta_+^2}-|\beta_+|) \nonumber
 \end{eqnarray}
were $\beta_{\pm}=\nu/[\sqrt{\pm(\sin^2\theta-\nu^2)}\cos \theta]$,

The GF $g_R^r$ in the main text (Eq.(14))  ready follows after  integration over $\theta$ variable: $g_d=\int_0^{\pi} d\theta G_d$ and $g_{off}=\int_0^{\pi} d\theta G_{off}$.
\noindent



\end{document}